\documentclass[letterpaper, 10 pt, conference]{ieeeconf}  

\IEEEoverridecommandlockouts                              

\usepackage{graphics} 
\usepackage{epsfig} 
\usepackage{amsmath} 
\usepackage{amssymb}  
\usepackage{url}
\usepackage{array}

\newcolumntype{x}[1]{%
&gt;{\raggedleft\hspace{0pt}}p{#1}}%

\newcommand{\distCoarse}{\Delta}
\newcommand{\distFine}{\delta}
\newcommand{\simplemean}{\mathit{mean}}
\newcommand{\simplevar}{\mathit{var}}

\title{\LARGE \bf
Measuring the impact of cognitive distractions on driving performance using time series analysis
}

\author{Matias Garcia-Constantino$^{1}$, Paolo Missier$^{1}$, Phil Blythe$^{2}$ and Amy Weihong Guo$^{2}$
\thanks{*This work was supported by the Social Inclusion through the Digital Economy (SiDE) project at Newcastle University.}
\thanks{$^{1}$Matias Garcia-Constantino and Paolo Missier are with the Computing Science Department, Newcastle University, United Kingdom.
        {\tt\small matias.garcia-constantino@ncl.ac.uk, paolo.missier@ncl.ac.uk}}%
\thanks{$^{2}$Phil Blythe and Amy Weihong Guo are with the Transport Operations Research Group, Newcastle University, United Kingdom.
        {\tt\small phil.blythe@ncl.ac.uk, weihong.guo@ncl.ac.uk}}%
}

\begin{document}

\maketitle
\thispagestyle{empty}
\pagestyle{empty}

\begin{abstract}
Using current sensing technology, a wealth of data on driving sessions is potentially available through a combination of vehicle sensors and drivers' physiology sensors (heart rate, breathing rate, skin temperature, etc.).
Our hypothesis is that it should be possible to exploit the combination of time series produced by such multiple sensors during a driving session, in order to (i) learn models of normal driving behaviour, and (ii) use such models to detect important and potentially dangerous deviations from the norm in real-time, and thus enable the generation of appropriate alerts.
Crucially, we believe that such models and interventions should and can be personalised and tailor-made for each individual driver.
As an initial step towards this goal, in this paper we present techniques for assessing the impact of cognitive distraction on drivers, based on simple time series analysis.
We have tested our method on a rich dataset of driving sessions, carried out in a professional simulator, involving a panel of volunteer drivers. Each session included a different type of cognitive distraction, and resulted in multiple time series from a variety of on-board sensors as well as sensors worn by the driver.
Crucially, each driver also recorded an initial session with no distractions. In our model, such initial session provides the baseline times series that make it possible to quantitatively assess driver performance under distraction conditions.
\end{abstract}

\section{Introduction}

The effect of distractions on drivers' performance has in the past been documented and quantified based either on accident reports (\cite{Klauer2006,Stevens2001,Stutts2001}), or on controlled experiments in a simulated driving environment, as discussed in Section \ref{sec:RelatedWork}.
While these studies are useful to inform policy and driving regulations, on-board systems for real-time alerts are also needed for accident prevention, based on real-time monitoring of drivers' response to a variety of distractions.
Furthermore, response to unexpected situations may vary greatly across the general drivers population, depending on experience, confidence levels, and current mental state and focus. It is therefore important that such systems be personalised for individual drivers.
The ultimate goal of our research is to help enhance road safety by enabling the creation of such personalised on-board systems,  based on models of drivers response to typical distractions.
To develop such models, we must first be able to quantify the impact of a variety of distractions on drivers' performance.
This is the goal of the preliminary research presented in this paper.
Specifically, we use sensor-based monitoring of both car and driver in a series of simulated driving scenarios, exercised on a panel of volunteer drivers, to understand the signals associated with a variety of realistic distractions, as well as the appropriate data processing techniques.

We focus on mental distractions, which involve an increase on the cognitive load of the driver away from the main task in non-emergency scenarios, typically in the form of conversations with passengers and/or with other parties over a phone (either hand-held or hands-free)\footnote{On the other hand, in our study we ignore visual and manual distractions, where the eyes or hands are used on other tasks while driving.}.
In contrast to studies that are based on accidents that were reportedly caused by documented distractions, such as usage of a mobile device, we are able to account for less drastic consequences, ranging from ignoring navigation system instructions, to the gradual increase in stress level as indicated by irregular driving patterns or physiology indicators (e.g. changes in heart rate).

%

Following Klauer \emph{et al.} \cite{Klauer2006}, we focus on the two most common types of distraction-inducing activities: (i) Passenger-Related Secondary Task (Passenger in rear seat\footnote{The case of having a passenger in the front seat was not considered due to the technical limitations of the driving simulator (i.e. lack of space for a front seat passenger).}) and (ii) Wireless Device (talking/listening using a mobile phone either hands-free or hand-held, locating/reaching/answering mobile phone).
Our experiments include variations of these activities, i.e. different types of phone conversations associated with different levels of cognitive load, which are carried out in a controlled, simulated and sensor-rich driving environment.

\begin{figure}[thpb]
 	\centering
 	\includegraphics[width=\columnwidth]{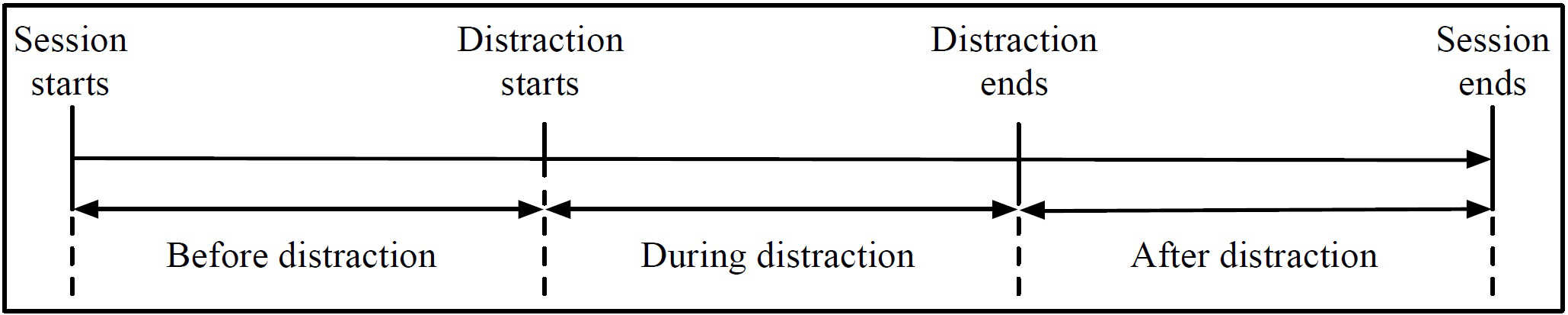}
   \caption{Segmenting driving sessions.}
   \label{fig:DrivingSessionDiagram}
 \end{figure}

More specifically, the research involves a panel of volunteers. 
Each volunteer is asked to take several test drives along a set course in a professional car simulator (of the type normally used to train candidates for a driving test).
During each driving session, the volunteer is subject to a different type of distraction. Unlike sudden and short-lived emergencies, such as for instance a pedestrian stepping onto the street, our experiments have a set duration, and they are made to start and end at the same place on the course for each volunteer.
 Thus, we characterise distractions by a ``before'' phase where ordinary driving occurs, a main event (e.g. the phone ringing), a ``during'' phase, and an ``after'' phase, marked for instance by the end of the phone call (Fig. \ref{fig:DrivingSessionDiagram}).
 Additionally, each volunteer provides a set of \textit{baseline} time series, captured during the course of a distraction-free driving session.
 Each session is therefore described by a collection of time series, each captured by a different car control from the driving simulator (braking, steering, acceleration, etc.) as well as a physiology signal (heart rate), and synchronised over a common clock.
 
Our main contribution is the use of time series analysis techniques, namely Dynamic Time Warping (DTW) and distance metrics, to compare time series from different phases of the driving sessions with a \textit{baseline} session, in which no distractions take place. The use of time series analysis keeps the fine grained details because the data is not summarised. In our design, such baseline plays the role of a model of ``normal'' driving behaviour, against which the behaviour in the experiment sessions is compared. Note that while drivers were not instructed to a target speed or a following distance, there were speed limits placed in the route which they followed. There were no further instructions in order to make the driving session closer to a real life one.
We use our growing collection of experimental datasets, collected from volunteers, to demonstrate the use and benefits of such techniques over simpler experiment designs that do not make use of baseline data.

%


\section{Related work}
\label{sec:RelatedWork}

The research literature on car drivers and specifically on behaviour induced by distractions is quite extensive. Here we focus specifically on the aspects that are most relevant to our own research, namely those addressing cognitive distractions such as the use of mobile phones and having conversations with passengers.

%
As background for the area, Young and Regan \cite{Young2007} provide a review of the literature related to driver distractions that focuses on the use of mobile phones, covering many areas of driver distractions, including (i) Secondary Task Demands, (ii) Driving Task Demands, (iii) Driver Age and Experience and (iv) Driver Distraction and Crash Risk. An interesting element is ``compensatory behaviour'', which is how drivers compensate the lack of attention due to the use of a mobile phone by modifying their driving in ways such as reducing speed, checking mirrors less frequently or not calculating the distance towards another vehicle.

Research into the effect of using mobile phones on driver's attention and performance began with the first generation of portable phones, in the early '90s.
In one such work, Brookhuis \emph{et al.} \cite{Brookhuis1991} used Heart Rate Variability (HRV) as a means to evaluate the mental workload of the drivers along with other factors such as lateral position, steering wheel movements and the frequency of checking the rear view mirror.
They concluded that the use of mobile phones decreased traffic safety and recommended the use of hands-free instead.
However, results based on HRV alone may be limited, as HRV changes sharply in response to emergencies, and more smoothly in the presence of physical stress, but not so much in the presence of long-lived distractions.
McKnight and McKnight \cite{McKnight1993} presented a study in which participants had to watch a video of a driving sequence and were expected to respond to traffic situations while being under five types of distractions, three of which involved the use of mobile phones. It was shown that the use of mobile phones affects driver attention regardless the age of the driver.

The effects of mobile phone conversations on the driving performance, using a driving simulator, were also investigated  by Strayer and Drews \cite{Strayer2004}, with focus on 
comparing performance between younger and older drivers.
While their results \cite{Strayer2004} indicate that the reactions of both younger and older drivers decreased while using mobile phones, interestingly the average reactions of younger drivers while using a mobile phone were similar to those of older drivers in the absence of such distractions.
Similarly, Shinar \emph{et al.} \cite{Shinar2005} used a simulator to compare reactions across age groups (young, middle age and old). Similar to our approach, they used multiple 
measures, including average speed, speed variance and steering wheel deviations.

More simulator-based results for hand-held and hands-free mobile phone conversations are available from Haigney \emph{et al.} \cite{Haigney2000}, using mean and standard deviation of heart rate, speed and variability of accelerator pedal travel before, during and after the call. Both groups of authors comment on compensatory behaviour of drivers while using a mobile phone.
Yet another study in this space, by Rakauskas \emph{et al.} \cite{Rakauskas2004}, also investigated the effects of mobile phone conversations on driving performance using a  simulator as well as a similar set of measures, namely accelerator variability, speed variability, average speed, steering offset, mean lateral speed, reaction time, collisions and mental workload. Once again, the conclusion is that the effects of mobile phone conversations while driving are higher workloads and a decreased driving performance.
Finally, the study by Drews \emph{et al.} \cite{Drews2008} is the closest to ours in terms of the situation and factors considered, i.e., the effects on driving performance of having passenger and mobile phones conversations, again assessed using a simulator. In this case, lane keeping, mean speed and mean distance were used for the assessment.

In contrast to the aforementioned methods used for investigating driver distractions, our method is based on time series alignment and comparison. As it will be shown in Section \ref{sec:Evaluation}, additionally to Euclidean Distance we use some of the evaluation measures used in other works for comparison purposes.

\section{Approach}
\label{sec:ProposedMethod}

As mentioned, a single driving session is described by a collection of time series, one for each available sensor and signal, i.e., speed, braking and steering activity, heart rate, etc. The signals used were considered because of their relevance and their previous use in related papers.
We are going to denote each session with $D_j$, where $j$ indicates one of a set of possible planned distractions. The $i$-th time series within the session is denoted $D_j[i]$. 
Importantly, a baseline session $D_0$ involving no distractions is recorded by each volunteer at the start of their driving experience.
As depicted in Fig. \ref{fig:DrivingSessionDiagram}, start and end-of-the-distraction events are associated with each $D_j$.
This determines a natural segmentation: before/during/after distraction on each $D_j[i]$. We denote such segments $D_j[i].b$, $D_j[i].d$, and $D_j[i].a$, respectively, and $D_j[i].s$ for a generic segment type $s$ within a series.

\subsection{Derived similarity series and segment distance}

Given two time series $R = [r_1 \dots r_n]$ (the \textit{reference} series) and $Q = [q_1 \dots q_n]$ (the \textit{query} series), we compute two measures of similarity between $R$ and $Q$.

\begin{itemize}
\item \textit{Coarse-grained} distance is defined as the Euclidean distance:
\begin{equation}
\distCoarse(Q,R) =  \sum_{j:i}^{n} \sqrt{(r_j - q_j)^2}
\label{eq:overallDist}
\end{equation}

\item \textit{Fine-grained} distance is itself a series, obtained by repeatedly computing the Euclidean distance over a sliding window of size $w$:
\begin{equation}
\distFine(Q,R,w) = [s_1 \dots s_l]
\label{eq:segmentDist}
\end{equation}
 is a series of length $l = n - w +1$, where:
\[ s_i =  1- \sum_{j:i}^{i+w-1} \sqrt{(r_j - q_j)^2}\]
\end{itemize}

In our assessment, we want to compare time series for each segment $s$ type in the experiment, $D_j[i].s$, with the corresponding segment type $D_0[i].s$ in the baseline, and across all series $i$. That is, for each distraction type $j$ we set $Q$ and $R$ as follows:
\begin{align*}
\mbox{ \textit{before}: } & Q = D_j[i].b, & R = D_0[i].b \\
\mbox{ \textit{during}: } & Q = D_j[i].d, & R = D_0[i].d \\
 \mbox{ \textit{after}: }  & Q = D_j[i].a, & R = D_0[i].a
\end{align*}
In addition, we are also going to compare series that correspond to different stretches of road within the ``during'' segments, i.e., straight road vs more challenging turns.
Our general expectation is that the \textit{before} distance to the baseline will be small compared to the \textit{during} distance. The \textit{after} measure may provide an indication of the lingering effect of the distraction on the rest of the driving session.

As mentioned, most past research on detecting the effect of distractions simply compares the ``before'' and ``during'' phases without regards to a specific, individually set baseline. 
In order to account for this simpler design, we also compute the mean and variance of the signal $D_j[i].s$ for each segment type $s$ and within each time series $D_j[i]$,  which we denote by $\simplemean(D_j[i].s)$, and $\simplevar(D_j[i].s)$, respectively.

\subsection{DTW alignment}

Although all sequences are recorded using the same route, they are bound to slightly vary in length. To account for these variations, we use Dynamic Time Warping (DTW) to align the sequences prior to computing their similarity.
DTW is a well-known general technique used to assess the extent of the match between a query sequence $Q$ within a reference sequence $R$, while accounting for differences in length and scale between $Q$ and $R$ \cite{Giorgino2009}.
DTW relies on two warping functions, $\phi_q$ and $\phi_r$, which are used to remap the time indices of $Q$ and $R$, resulting in a non-linear temporal alignment. The optimal alignment $\phi$ is given by the minimum global dissimilarity, or DTW distance:
\[ D(Q, R) = min \; d_\phi(Q, R)  \]

\begin{figure}[thpb]
	\centering
	\framebox{\parbox{.7\columnwidth}{\includegraphics [width=.7\columnwidth]{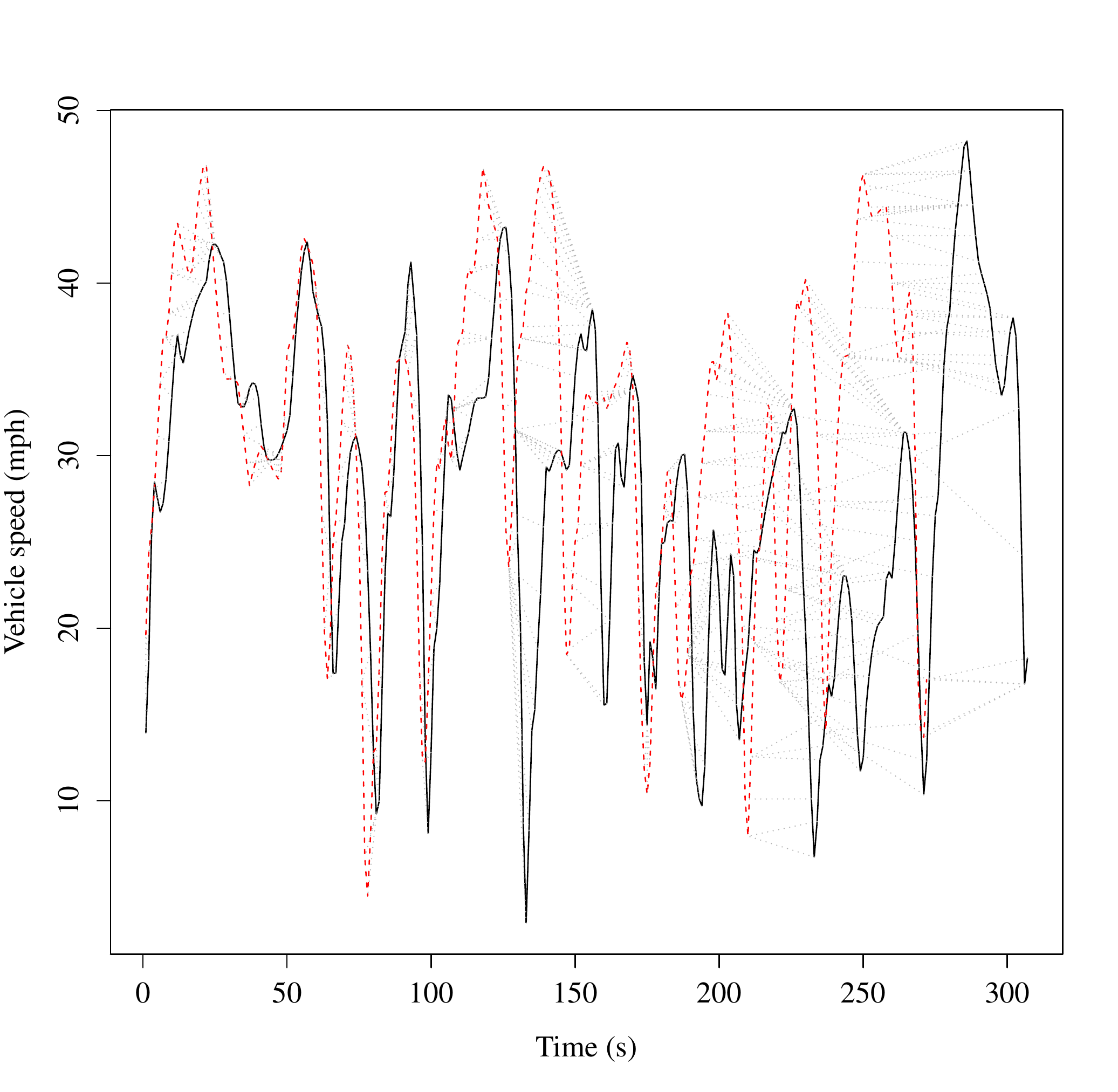}}}
  \caption{Vehicle speed time series DTW alignment.}
  \label{fig:VehicleSpeedTimeSeriesDTWAlignment}
\end{figure}
\begin{figure}[thpb]
	\centering
	\framebox{\parbox{.7\columnwidth}{\includegraphics[width=.7\columnwidth]{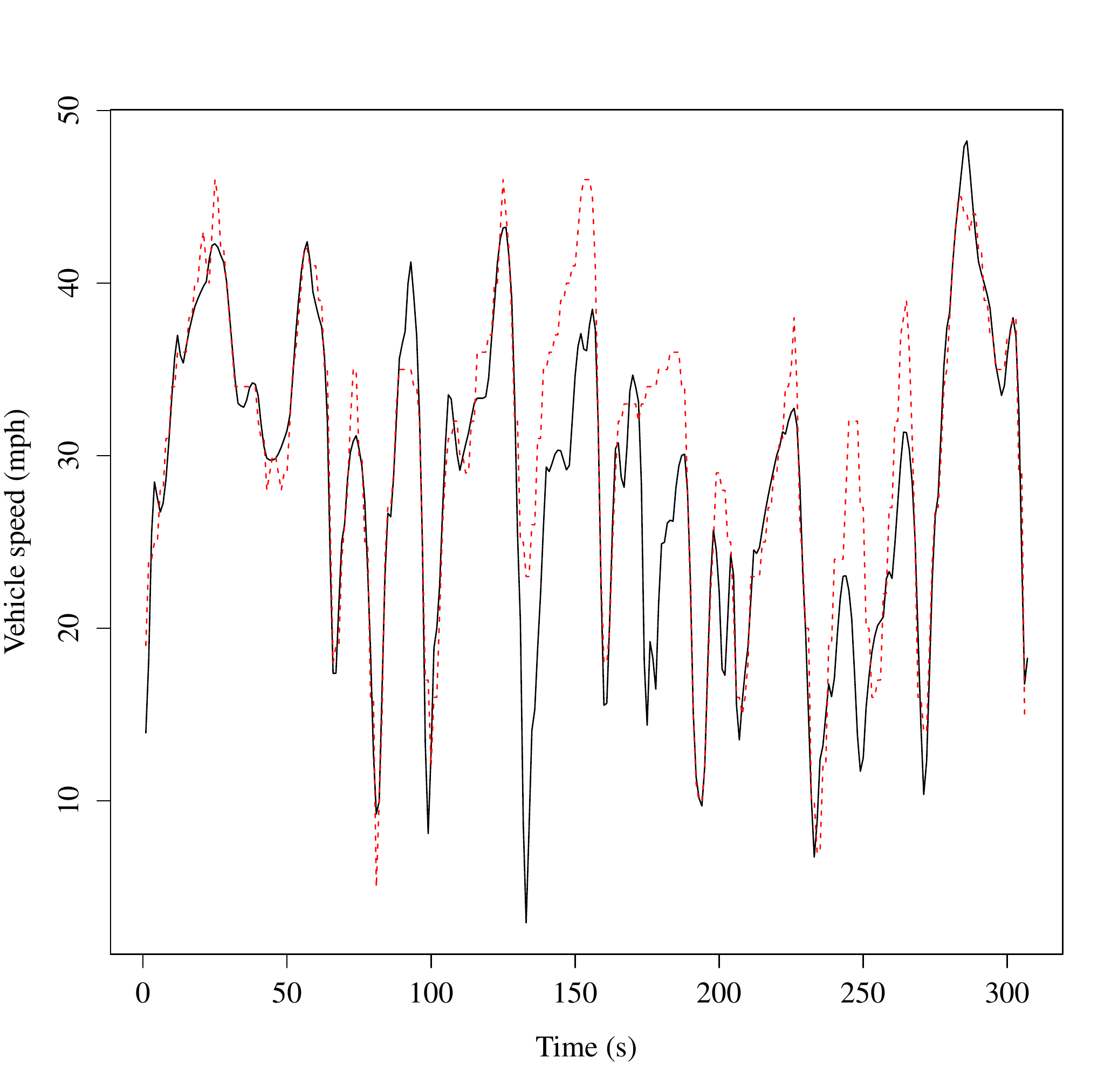}}}	
  \caption{Vehicle speed time series aligned.}
  \label{fig:VehicleSpeedTimeSeriesAligned}
\end{figure}

As an example consider Fig. \ref{fig:VehicleSpeedTimeSeriesDTWAlignment}, where $Q$ (solid black) and $R$ (dashed red) are two series representing vehicle speed, for a distraction session and for the baseline session, respectively.
DTW transforms $Q$ into the new sequence shown in Fig. \ref{fig:VehicleSpeedTimeSeriesAligned} (again dashed red).

%

\subsection{Computing similarity plots}

Using DTW in combination with the similarity function, for each driving session $D_j$ we process each sequence $D_j[i]$ in $D_j$ as follows.

\begin{enumerate}
\item realign using DTW to align the query $Q = D_j[i]$ to the reference $R = D_0[i]$. This results in a new query sequence, $Q'$;

\item split both $Q'$ and $R$ into their before/during/after segments, using the distraction start and end timestamps recorded for $D_j$;

\item compute:
\begin{align*}
&\distFine(Q'.b,R.b,w), &\distCoarse(Q'.b, R.b)\\
&\distFine(Q'.d,R.d,w), &\distCoarse(Q'.d, R.d)\\
&\distFine(Q'.a,R.a,w), &\distCoarse(Q'.a, R.a)\\
\end{align*}
A window size $w = 10$ was used in all cases.
\end{enumerate}

This process produces fine-grained distance plots for each series $D_j[i]$ and for each $D_j$, as well as overall, coarse-grained distance measures for each experiment segment from its baseline.
By applying this processing over all monitoring variables and all types of distractions, and across all volunteers in our panel, we obtain a rich collection of distance measures.
In the next section we analyse these measures.
We emphasise again here that in this paper we use these figures only as an example to illustrate the benefit of using personalised ``normal behaviour'' time series for a variety of observable driving variables, in combination with experiments on non-emergency distractions, as a way to enable individualised drivers' assessment.

\section{Evaluation}
\label{sec:Evaluation}

\subsection{Experiment design}
\label{sec:ExperimentsDevicesSettingsDataCollectionAndPreprocessing}

\begin{figure}[thpb]
	\centering
	\framebox{\parbox{3.2in}{\includegraphics[scale=0.22]{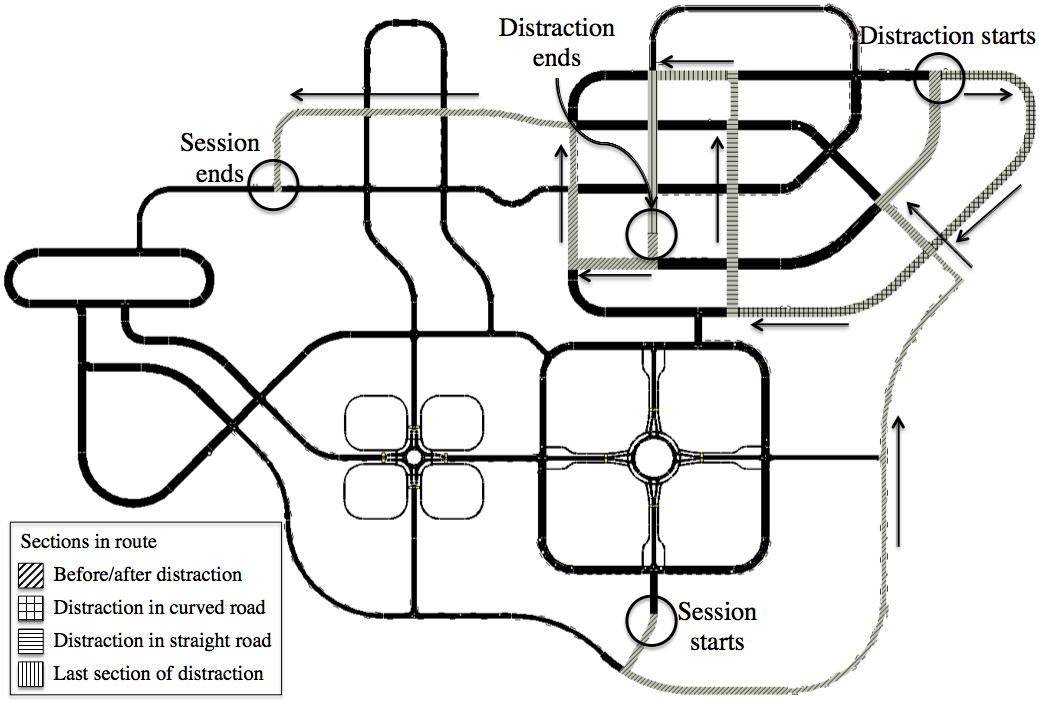}}}
  \caption{Route map.}
  \label{fig:RouteMap}
\end{figure}

The experiments involved a panel of 16 volunteer drivers (10 men and 6 women) aged between 20 and 50 (half of them in the range of 17-25 and the other half in the range of 26-50).\footnote{
A higher number of participants took part, however a number of them experienced simulator sickness, a recurrent issue in experiments that involve driving simulators, which prevented them from completing the experiments. Data from those participants was discarded. Brooks \emph{et al.} \cite{Brooks2010} presented a comprehensive study on simulator sickness during driving simulator studies where the relationship between motion sickness and simulator sickness is discussed.} Although the analysis presented does not consider drivers' age and gender, the distribution was intended to be as even as possible. The range of driving experience of the volunteers was between 5 and 20 years, with an average of 8.85 years.
Each participant was asked to undergo a sequence of five driving sessions in a driving simulator, with short breaks in between.
All sessions followed the same route, shown in the map of Fig. \ref{fig:RouteMap}), with the help of a (simulated) sat nav system with vocal instructions.
The map shows the three main route segments, namely before/during/after distraction (Fig. \ref{fig:DrivingSessionDiagram}), as well as different features within the distraction segment (a curve, a straight stretch).
Drivers were only asked to follow road rules, but they were otherwise not given specific instructions on driving style (e.g. target speed, distance from other vehicles).
This was done deliberately, to simulate a realistic noisy scenario, which includes randomly generated surrounding traffic (which followed traffic rules).

Each session included a different type of distraction-inducing tasks, designed to increase cognitive load away from the driving task.
These are summarised in Table \ref{fig:DrivingSessionsAndCognitiveDistractions}.
The tasks for the first three driving sessions (DS1, DS2, DS3) included a phone caller / simulated back seat passenger asking the driver to recall information from films, books or TV shows over a real phone call (both hands-free and hand-held).
The session indicated as DS4 in the table is the baseline session, denoted $D_0$ in the previous section.
The task in the last session (DS5) included answering questions about basic mathematics operations and spelling words.

\begin{table}[htp]
\centering
\caption{Driving sessions and associated cognitive load.}
\label{fig:DrivingSessionsAndCognitiveDistractions}
\begin{tabular}{|c|c|c|} \hline
\bf{Driving} & \bf{Cognitive distraction} & \bf{Type of conversation} \\ 
\bf{Session} &                            &               \\ \hline             
DS1          & Passenger in the back seat & Recalling information \\ \hline
DS2          & Mobile phone hands-free    & Recalling information \\ \hline
DS3          & Mobile phone hand-held     & Recalling information \\ \hline
DS4          & No distractions            & No conversation \\ \hline
DS5          & Passenger in the back seat & Maths \& Spelling \\ \hline
\end{tabular}
\end{table}

\begin{figure}[thpb]
	\centering
	\framebox{\parbox{\columnwidth}{\includegraphics[width=\columnwidth]{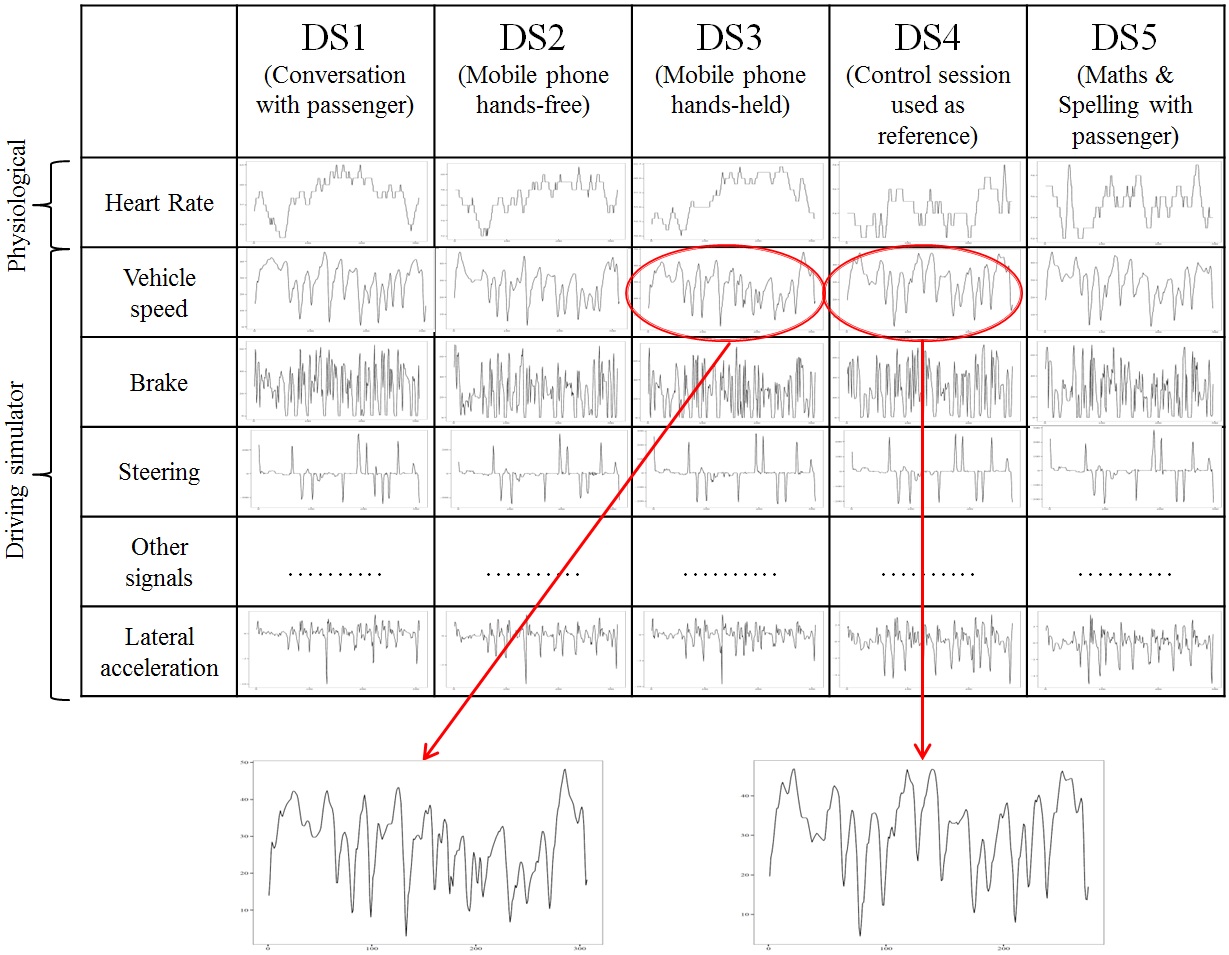}}}
  \caption{Time series collected from sensor variables are collected for multiple driving sessions, with distractions (DS1, DS2, DS3, DS5). DS4 is the baseline session.}
  \label{fig:DrivingSessionsAndRelevantSignals}
\end{figure}

After each session, data pre-processing included (i) outlier elimination, general cleaning, ensuring proper formatting, and (ii) time-synchronizing the physiology data with the simulator data. 
Fig. \ref{fig:DrivingSessionsAndRelevantSignals} summarises the driving sessions and series collection phase. Columns denote the different driving sessions $D_j$, rows denote the monitoring variables $i$, and individual cells represent the series within each session, $D_j[i]$.
     
\subsection{Equipment}

All experiments were conducted using a professional driving simulator, commonly used in driving schools, available at the Intelligent Transport lab at Newcastle University.\footnote{
The lab is funded by the SiDE / inclusive transport project (Social Inclusion through the Digital Economy, \url{http://www.side.ac.uk/inclusive-transport}.
 The simulator is manufactured by ST Software Simulator Systems \url{http://www.stsoftware.nl/}}
Amongst the available car sensors in the simulator, the following were used:  vehicle speed, gear change, brake, accelerator, clutch, steering, lateral acceleration, longitudinal acceleration, RPM. Road position, lane position and path along the route are also available in all cases.
In addition to car data from the simulator, we also collect basic physiology data by fitting drivers with a commercial bioharness with heart rate monitoring capability, commonly used for fitness applications.\footnote{Zephyr:  \url{http://www.zephyranywhere.com/products/}.}

\subsection{Approach to analysis}
\label{sec:OverviewOfTheObtainedResults}

The main conjecture we want to support, using our dataset, is that using a general population model as a reference to assess the impact of non-emergency distractions is insufficient, because differences in individual drivers behaviour require personalised models for a fair assessment.
Instead, we suggest the need for personalised models of driving behaviour, as a basis to develop effective interventions to promote safety.

Thus, rather than simply comparing sensor values for the ``before'' and the ``during'' segments across a population of drivers, we offset individual variability by using the distances between before/during segment within an experiment, and the corresponding baseline segments for the same driver (which represents the driver's ``normal'' behaviour).
For instance, it is entirely possible that heart rate at the start of a distraction session is higher than it was at the start of the baseline session, for a number of reasons.
By considering the distance between the before segments within the experiment and in the baseline, we account for this systematic difference
\footnote{Note that due to space constraints, the discussion below is limited to only a few types of series (two rows in Fig. \ref{fig:DrivingSessionsAndRelevantSignals}), primarily heart rate and vehicle speed, which have been used in prior literature to evaluate the effect of drivers distractions (\cite{Young2007, Brookhuis1991, Shinar2005, Haigney2000, Rakauskas2004, Drews2008}).}.

\subsection{Analysis by individual coarse-distances relative to the baseline}

We follow this idea in our first analysis. Here for each driver, for each type of distraction experiment $j$ and each sensor variable $i$, we compute the set of coarse-grained distances for ``before'' segments and for ``during'' segments, as follows:
\[ \Delta_b(j,i) = \distCoarse(D_j[i].b, D_0[i].b)\]
\[ \Delta_d(j,i) = \distCoarse(D_j[i].d, D_0[i].d)\]
and we report their relative difference:
\[ \Delta_{b,d}(j,i) = \frac{| \Delta_b(j,i) - \Delta_d(j,i) |}{\Delta_b(j,i)}  \]
These figures provide an overall indication of the amplitude of the effect of the distraction for a given sensor variable, when such effect is assessed against each driver's baseline.
Tables \ref{fig:deltas-detail} and \ref{fig:deltas-summary} show sample results computed for selected sensor types and distraction types.
Space constraints prevent us from reporting additional relative distance figures, e.g. to compare straight vs turn segments within a distraction segment.
Still, as a general observation we note that all sensor variables shown provide detectable relative differences between before and during segments for each participant. 
However, the important point here is that such distances vary widely across participants.
This observation supports our suggestion that in order to deploy effective interventions to ensure drivers' safety, individual, personalised models of driving behaviour are required.


\begin{table}[htp]
\centering
\caption{Examples of relative distances $\Delta_{b,d}(j,i)$ where $i$ is Heart Rate (HR), Vehicle Speed (VS) and Brake, and $j$ is distraction type DS1. Distances are significant and easy to detect across all participants relative to their own baselines, but exhibit high variability across participants.}
\label{fig:deltas-detail}
\begin{tabular}{|c||c|c|c|} \hline
Distraction type & \multicolumn{3}{c|}{} \\
DS1 & \multicolumn{3}{c|}{Average reading per sensor variable ($\%$)} \\ \hline
Participant & HR & Vehicle Speed (VS) & Brake \\ \hline
P1  &  10 &   2 &  16 \\
P2  &  60 &  11 &  14 \\
P3  &  14 &  10 &  13 \\
P4  &   3 &  68 &  20 \\
P5  & 142 &  12 &   6 \\
P6  &  23 & 130 &  27 \\
P7  &   8 &  46 &  18 \\
P8  &   5 &  31 &  11 \\
P9  &  25 & 264 &  66 \\
P10 & 187 & 224 &   0 \\
P11 &  17 &  19 &   1 \\
P12 & 165 &  17 &  12 \\
P13 &  17 &  25 &  29 \\
P14 &  54 &   9 &   2 \\
P15 &  62 &  58 &   9 \\
P16 &  45 &  45 &   9 \\ \hline
avg &  52 & 61 & 16 \\
stddev & 35 & 62 & 2 \\ \hline
\end{tabular}
\end{table}


\begin{table}[htp]
\centering
\caption{Summary of relative distances $\Delta_{b,d}(j,i)$ for all distraction types and three sensor variables, across participants.}
\label{fig:deltas-summary}
\begin{tabular}{|c| |c|c| |c|c| |c|c|} \hline
 & \multicolumn{2}{c||}{HR ($\%$)} & \multicolumn{2}{c||}{VS ($\%$)} & \multicolumn{2}{c|}{Brake ($\%$)} \\ \hline
Distraction type & mean & var & mean & var & mean & var \\ \hline
DS1 & 52 & 35 & 61 & 62 & 16 & 2 \\
DS2 & 44 & 19 & 52 & 32 & 15 & 2 \\
DS3 & 92 & 165 & 53 & 55 & 20 & 2 \\
DS5 & 75 & 92 & 70 & 20 & 24 & 4 \\ \hline
\end{tabular}
\end{table}

%

\subsection{Analysis by paired designs}

Next, we consider the statistical significance of sensor variables readings to quantify the effect of distractions. 
As reported in Section \ref{sec:RelatedWork}, most studies are based on a standard before/during design, where participants are asked to follow driving guidelines to limit noise in the observations.
In contrast, our experiments are closer to the driving ``in the wild'', as no such guidelines were issued.
We can however reproduce such designs using our data, by simply considering the mean values of sensor variables for the before and during segments and for each distraction, ignoring the baseline sessions and thus time series distances altogether.
This leads to a set of standard paired experiment designs, one for each combination of distraction type and sensor type.
More precisely, given a distraction session $D_j$ and a sensor variable $i$, we build two sample distributions, for $D_j[i].b$ and $D_j[i].d$, where each participant contributes one 
before/during sample pair to each of the two distributions.
Since each sample can be shown to be approximately normally distributed\footnote{We test for normality using a standard QQ plot.}, we can test for the difference of the means of the distributions using a standard paired t-test. 
The results, reported in Table \ref{fig:p-values-means}, show systematically small p-values (the RPM column has been chosen as one of the few exceptions), suggesting that all sensor variables can potentially be used (one at a time) to measure impact across all distractions. 
Note that covariate analysis has not been performed on these figures, as it is left for further work.

\begin{table}[htp]
\centering
\caption{p-values for a paired t-test on the mean values for before/during segments on four sensor variables (16 samples).}
\label{fig:p-values-means}
\begin{tabular}{|c|c|c|c|c|} \hline
Distraction type & HR & Brake & VS & RPM \\ \hline
DS1 & 2E-02 & 2E-04 & 9E-06 & 0.24 \\
DS2 & 5E-03 & 1E-04 & 1E-06 & 0.02 \\
DS3 & 6E-05 & 3E-05 & 4E-05 & 0.80 \\
DS5 & 1E-04 & 1E-04 & 1E-05 & 0.20 \\ \hline
\end{tabular}
\end{table}


These results do not account for baseline series, and thus do not account for individual variability amongst drivers.
When time series are measured using distances from each participant's own baseline, the results are much less clear-cut. 
To study this case, we use an alternative design, in which the simple mean values of sensor variables are replaced with the distances of the experiment time series from their own corresponding baselines.
The results, for the same sensor variables used in Table \ref{fig:p-values-means}, are shown in Table \ref{fig:p-values-distances}.
Note that in this instance a two-tailed nonparametric test (Wilcoxon signed rank) was used as using distances, the samples are not necessarily normally distributed. However a paired t-test produces fairly similar results.

\begin{table*}[htp]
\centering
\caption{p-values for before/during segments when the samples consist of time series distances from the baseline (Wilcoxon signed rank test). Significant p-values are highlighted.}
\label{fig:p-values-distances}
\begin{tabular}{|c|c|c|c|c|c|c|c|c|c|c|c|} \hline
Distraction type & HR & Gear & Brake & Accelerator & Clutch & Steering & AccLat & AccLong & LanePos & VS & RPM \\ \hline
DS1 & 0.90 & 0.50 & 0.80 & 0.20 & 0.70 & 0.20 & 0.60 & \bf{0.03} & 0.40 & 0.70 & 0.40 \\
DS2 & 0.50 & 0.80 & \bf{0.08} & \bf{0.06} & 0.50 & \bf{0.04} & 0.17 & 0.63 & 0.25 & 0.07 & 0.60 \\
DS3 & \bf{0.08} & 0.70 & 0.50 & 0.90 & 0.50 & 0.97 & 1.00 & 0.63 & \bf{0.002} & 0.50 & 0.20 \\
DS5 & \bf{0.03} & \bf{0.06} & \bf{0.05} & \bf{0.05} & 0.20 & 0.20 & \bf{0.004} & 0.56 & \bf{0.04} & \bf{0.003} & \bf{0.02} \\ \hline
\end{tabular}
\end{table*}


As we can see using this method, significant difference between ``before'' and ``during'' are now much more sporadic.
We believe this type of test provides a more realistic setting than the standard design, resulting in more stringent requirements on the sensor variables used for assessment. 
No single sensor variable seems adequate to account for before/during differences, indicating the need for further analysis.
 
\subsection{Role of fine-grained distance series}

\begin{figure}[tb]
\centering
\includegraphics[width=\columnwidth]{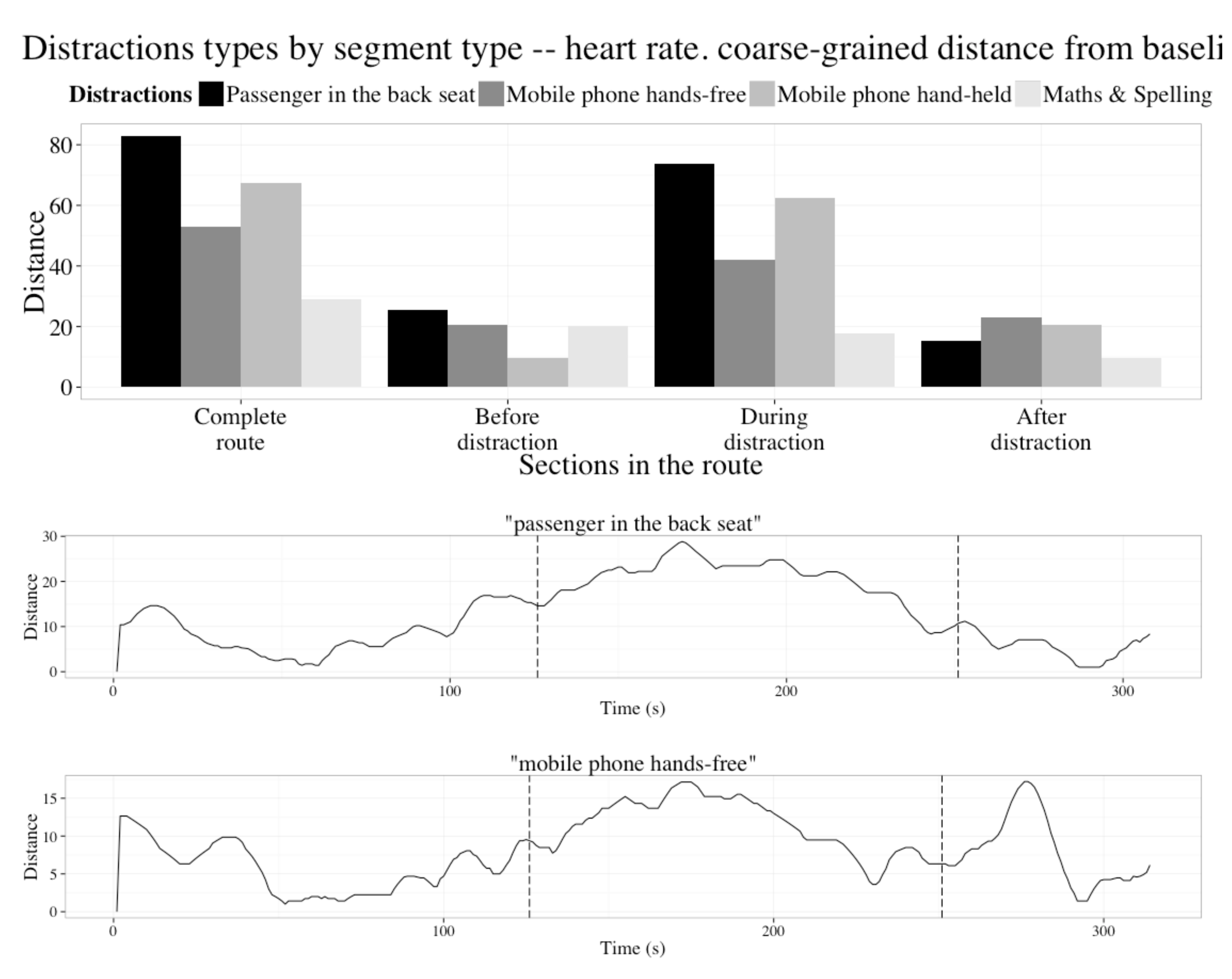}
\caption{Coarse and fine-grained distances for heart rate, participant P1.}
\label{fig:Participant1_heartRatePlotSummary}
\end{figure}

\begin{figure}[tb]
\centering
\includegraphics[width=\columnwidth]{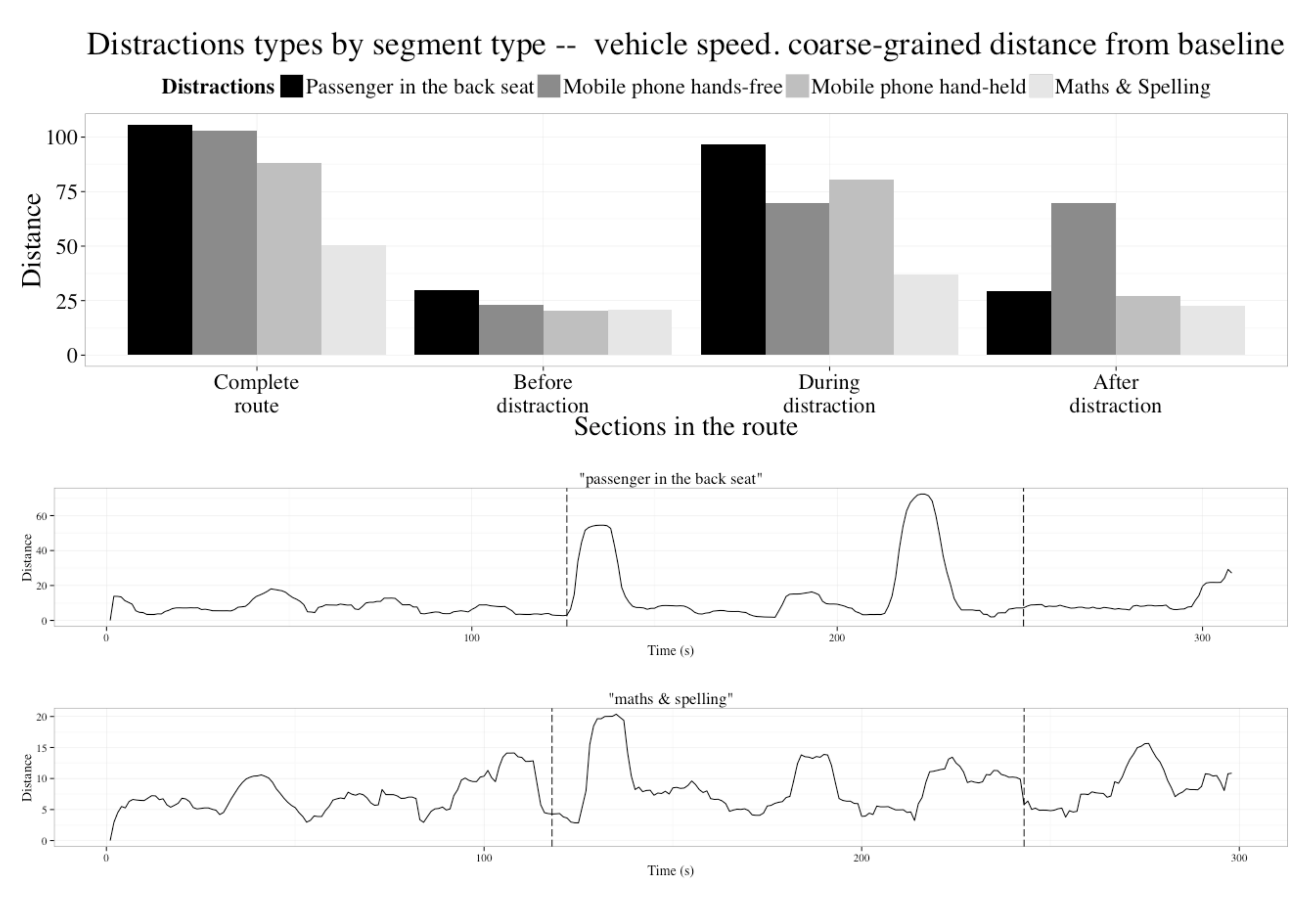}
\caption{Coarse and fine-grained distances for Vehicle speed, participant P1.}
\label{fig:Participant1_VSPlotSummary}
\end{figure}

Coarse-grained distance between two time series average the distances between corresponding points in the two series (after DTW alignment).
In contrast, computing the distance between two series over a sliding window, as defined in Eq. \ref{eq:segmentDist} provides a more precise characterisation of the points within the segment, where distance increases (note that distance is an absolute value, and is manifested as a spike in the distance series regardless of sign. For instance, drivers are known to compensate for distractions by reducing their speed).
Again owing to space constraints, here we only show two examples (Fig. \ref{fig:Participant1_heartRatePlotSummary} and \ref{fig:Participant1_VSPlotSummary}) for heart rate and vehicle speed for the same driver, where coarse-grained and fine-grained distances are combined.

\section{Summary and Conclusions}
\label{sec:ConclusionsAndFutureWork}

Our overall research goal is to develop models of drivers' behaviour, which can be used to monitor the performance of drivers and issue appropriate warnings, in a way that is personalised and aware of the individual driving style.
In this paper we have proposed a method for quantitative assessment of the impact of cognitive distractions on drivers, which is a pre-requisite to achieving this goal.
We start from two assumptions. Firstly, we assume that driving sessions are monitored using multiple sensors, for both car control and driver physiology.
Secondly, we assume that one or more traces of ``normal'' driving behaviour are available to be used as a \textit{baseline} to which actual observed behaviour can be compared.
To satisfy these assumptions, we have designed experiments in which 16 volunteer participants are asked to drive a close circuit in a car simulator, producing a baseline session and a set of experiment sessions, one for each different type of cognitive distraction. Each session consists of a collection of time series, one for each sensor variable.

We have defined both a \textit{coarse-grained} and a \textit{fine-grained} measure of distance between segments within a driving session, based on a combination of Dynamic Time Warping and Euclidean distance between time series. We have shown how these measures can be used to carry out a detailed analysis of the significance of individual sensor variables to explain the impact of a particular distraction.

We have tested our method on a variety of types of simulated cognitive distractions, namely: (i) conversation with a passenger in the back seat, (ii) conversation while using a mobile phone hands-free, (iii) conversation using a mobile phone hand-held and (iv) answering questions related to maths \& spelling. 

We believe this study can be instrumental to our main goal of learning personalised models of drivers' performance. 
Ultimately, we hope to be able to show that such models are more accurate than generic population-wide models, for the purpose of automatically generating effective alerts and interventions to promote driver safety on the road.

\addtolength{\textheight}{-12cm}   





\section*{ACKNOWLEDGMENTS}

This work was supported by the Research Councils UK Digital Economy Programme [grant number EP/G066019/1 - SIDE: Social Inclusion through the Digital Economy].


\bibliographystyle{IEEEtran}
\bibliography{IEEEabrv,DriverDistractions}

\end{document}